\documentclass[preprint]{aastex}

\newcommand{\iint}{\int\!\!\int}
\def\<#1>{\langle#1\rangle}
\def\pcite#1{[ref]}

\slugcomment{Revised: Mar 13, 2001}

\shortauthors{D. Chakrabarty \& P. Saha}
\shorttitle{Galactic Center Black Hole}

\begin{document}

\title{A Non-Parametric Estimate\\
of the Mass of the Central Black Hole\\
in the Galaxy}

\author{Dalia Chakrabarty\altaffilmark{1}}
\affil{Department of Physics (Astrophysics)\\
University of Oxford\\
Keble Road, Oxford OX1~3RH, UK}
\altaffiltext{1}{Department of Physics and Astronomy\\
University of Leicester\\
University Road, Leicester LE1~3RH, UK}
\and

\author{Prasenjit Saha}
\affil{Astronomy Unit, School of Mathematical Sciences\\
Queen Mary and Westfield College\\
London E1~4NS, UK}

\begin{abstract}
We estimate the mass of the central black hole in our Galaxy from
stellar kinematical data published by \citet{ghez} and
\citet{gerhard}.  For this we develop a method, related to
\citet{david}, for non-parametrically reconstructing the mass profile
and the stellar distribution function in the central region of the
Galaxy from discrete kinematic data, including velocity errors.
Models are searched using the Metropolis algorithm. We assume that the
mass distribution is spherical and the stellar velocity distribution
is isotropic, and devise a test of this assumption.  From proper
motions we obtain an enclosed mass of $2.0\pm{0.7}\times10^6{\rm
M}_{\odot}$ within the inner $0.0044\rm pc$, from radial velocities we
obtain a mass of $2.2^{+1.6}_{-1.0}\times10^6{\rm M}_{\odot}$ within
$0.046$pc and from three-dimensional velocities we obtain
$1.8^{+0.4}_{-0.3}\times10^6{\rm M}_{\odot}$ within $0.046$pc.
\end{abstract}

\keywords{Galaxy: center; Galaxy: kinematics and dynamics}

\section{Introduction}

The center of our Galaxy has been extensively studied at wavelengths
between the near infrared and radio regions in the electromagnetic
spectrum, \citep{sandqvist}. The dynamical center of our Galaxy is
considered to be coincident with a compact synchrotron radio source,
Sgr~A$^*$.  Sgr~A$^*$ is associated with a black hole of mass of the
order of 10$^6{\rm M}_{\odot}$, \citep{bower}. Kinematics in the
neighborhood of this black hole would provide a direct estimate of the
mass interior to the orbit. With this in mind, observational studies
of gas and stellar motions in the very central regions of our Galaxy
have been undertaken over the last few years.

This paper develops a new non-parametric method for mapping the
distribution function and the underlying potential of a spherical
isotropic stellar system, from discrete kinematic data, and applies it
to published radial velocity and proper motion data in the Galactic
center region to re-estimate the mass of the central black hole.

\section{Observational studies}

The notion that our own Galaxy could harbor a massive black hole at
its center was alluded to by theories that suggested the presence of
central massive black holes in galactic nuclei. Such a
hypothesis regarding the Milky Way gained stronger footing with the
availability of favorable observational evidence, starting with the
report on Ne II $12.8\mu$m emitting high velocity ionized gas at
the Galactic center by \citet{wollman}. An upper limit of
$4\times10^6{\rm M}_{\odot}$ was imposed on the mass enclosed within a
radius of $0.8$pc from the Galactic center. Similar observations of
the $12.8\mu$m Ne II emission from the Galactic center were
reported in subsequent years in \citet{lacy1} and
\citet{lacy2}. Later it was realized that the emission was from
large-scale flows of ionized gas, \citep{lo,serabynlacy,genzel85,gusten}.
The streamer velocity and position
could be fitted by orbital motions. Such fits indicated the presence
of a mass of $\approx{3}\times10^6{\rm M}_{\odot}$ within the inner
$0.5$pc.

However stellar motions were still needed to get a more reliable
estimate of the mass in the central regions in the Galaxy since gas
dynamics can be dictated by factors other than gravity.  Observations
of the early type stars that form a smaller cluster found in the
central $10''$ have also been made. Such an observational programme
led \citet{krabbe95} to predict the existence of a central mass of
$2-4\times10^6{\rm M}_{\odot}$ within the central $0.15$pc.
\citet{genzel1} reported the results of a radial motion study of $222$
stars, within the central 1pc of the dynamical center of our Galaxy.
This study predicted a central core of radius $\leqslant{0.06}$pc and
mass 2.2--3.2$\times10^6{\rm M}_{\odot}$.  The work done in
\citet{genzel2}, was the first study of proper motions of stars in the
Galactic center. Combining the radial motion data of \citet{genzel1}
and the proper motion data from \citet{genzel2}, Genzel and Eckart
reached the conclusion that the central core of our Galaxy has a mass
of $2.45\pm{0.4}\times10^6{\rm M}_{\odot}$ and lies within $0.015$pc
of Sgr~A$^*$. \citet{ghez} estimated the mass of the central dark core
of the Milky Way to be $2.6\pm{0.2}\times10^6{\rm M}_{\odot}$ and its
volume to be $\leqslant10^{-6}$pc$^3$. This conclusion was reached on
the basis of the proper motion data of 90 stars in their sample. The
latest report of stellar dynamics in the Galactic center is in
\citet{gerhard}.  In this analysis, anisotropy of stellar motions is
considered for the first time. The authors conclude that while
isotropy is a good assumption, there are distinct groups of stars that
display significant anisotropy in their motions. They estimate a
central mass of $2.6-3.3\times10^6{\rm M}_{\odot}$ depending on the
modeling of their data.

\subsection{Radial and proper motion study by Genzel et al.\ 
and Eckart \& Genzel} 

\citet{genzel1} observed radial velocities of 223 stars within
$\approx0.1$pc.  \citet{genzel2} reported proper motions of stars in
the central gravitational field of the Galaxy. The observational programme
spanned over a total of four years, at a resolution of $\approx0.15''$.
It was found that the velocity dispersions along the three mutually
orthogonal spatial axes were very similar, implying a very low degree
of central anisotropy.

A strong Keplerian fall-off of velocity dispersion with projected
radius [$\sigma(r_p)\propto r_p^{-0.5}$] was established, thus
strongly implying the existence of a central black hole.  This
provided a strong physical motivation to model the center as a point
mass and the velocity field as isotropic. The mass estimate of the
central core of the Galaxy could then be found by fitting the
parameters of the model to the radial and proper motion data.  A
simple virial estimate was supplemented by a \citet{BT} mass estimate
for the Galactic core. For this latter scheme the point mass was
considered to be embedded in an isotropic cluster of dispersions of
the order of $50$km/s at infinity. Virial and Bahcall-Tremaine mass
estimators gave similar values ($2.12\pm{0.7}\times{10}^{6}{\rm
M}_{\odot}$) for the mass within a projected radius of $0.24$pc. Using
the Jeans equations with the radial and proper motion data from
\citet{genzel1} and \citet{genzel2} implied the existence of a central
mass of $2.4\times10^6{\rm M}_{\odot}$ with a total (systematic and
fitting) uncertainty of about $\pm0.64\times10^6{\rm M}_{\odot}$. This
mass is considered to be concentrated within a radius of
0.015$\pm$0.02pc east of Sgr~A$^*$.

\subsection{Proper motion study by Ghez et al.}

In this work Ghez and co-workers observed proper motions of the
central cluster of the Galaxy. They followed the time evolution of the
positions of $90$ stars over a period of two years at a resolution of
$\approx0.05''$. The radial velocities were measured for stars at the
largest radii only, ($\approx0.1$pc). These few values of the
line-of-sight velocity dispersion, when compared to the projected
radial and tangential velocity dispersions indicated a good degree of
isotropy in the central regions.  Modeling the center as a point mass,
they estimated a central mass of $(2.6\pm{0.2})\times10^6{\rm
M}_{\odot}$ interior to a radius of $0.015$pc.

\subsection{Anisotropy incorporated analysis of stellar dynamics by
Genzel et al.}  In \citet{gerhard}, the data set used includes line of sight
velocity measurements reported earlier in \citet{genzel1}, improved values
of proper motions reported in \citet{genzel2}, and proper motions reported in
\citet{ghez}.  This ``homogenized best'' data set includes proper motions
of $104$ stars, with their errors as well as 227 radial velocities and
errors in these line-of-sight (LOS) velocities. For 32 stars both sky and
line-of-sight velocities are reported. The assumption of isotropy is
concluded to be broadly good, though motions of individual groups of
stars are spotted to show significant anisotropy, (examples of both
radial and tangential anisotropy have been spotted). The whole sample
of stars is divided into $7$ annuli and the projected radial and
tangential velocity dispersions corresponding to each annulus are
calculated. The global expectations of these projected velocity
dispersions can be related to the anisotropy parameter, $\beta$.  Thus
for each radial annulus, the anisotropy parameter is obtained. The
significance of the observed anisotropy is checked against data
generated by Monte Carlo clusters defined by simple anisotropic
distribution functions. Using the value of $\beta$ for each
annulus, \citet{LM}, Bahcall-Tremaine and virial mass estimates are
made for the very central region, the respective results being
$2.9\pm0.4\times10^6{\rm M}_{\odot}$, $3.1\pm0.3\times10^6{\rm
M}_{\odot}$ and $2.5-2.6\times10^6{\rm M}_{\odot}$.

\section{Motivation for a non-parametric scheme}
\label{sec:motivation}

The observational works discussed above obtain the mass
in the central region of the Galaxy using either the Jeans equation or
projected mass estimators like Bahcall-Tremaine, Leonard-Merritt, or
the virial theorem. The shortcomings of these methods are as follows.
\begin{itemize}
\item The implementation of Jeans equation requires smooth
approximations to the projected dispersions and the surface density,
that can be subsequently related to the total dispersion and the
spatial density via the Abel integral equation.  This suffers
from the problem of shot noise in radial bins, which tends to make the
Abel inversion unstable \citep{genzel1}. In addition, the Jeans
equations do not constrain the distribution function to be
non-negative.
\item The mass estimators are derived by taking moments of the Jeans
equations, and relate $\langle r^nM(r)\rangle$ (where $M(r)$ is the
enclosed mass, $n$ is an integer, and the average is over the stellar
density) to observed velocity dispersions.  Being moments, they
contain only limited information about the mass distribution.
Moreover, for $n=0$ (Bahcall-Tremaine and Leonard-Merritt estimators)
stars far from the center get more weight, which is undesirable when
estimating a black hole mass. Also, kinematic data sets in practice
sample only part of the potential, so some modeling or extrapolation
is necessary.
\end{itemize}

A non-parametric method, which can reconstruct both density and
distribution function without assuming functional forms for either,
would avoid all these problems.

\section{Practical algorithms used to solve the spherical inverse problem}
\label{sec:algorithm}

Here we present a scheme for studying the spherical inverse problem
under the assumption of isotropy. This algorithm is inspired by the
work in \citet{david}, but differs in some important ways: it uses
both radial velocity and proper motion data rather than just the
former, and uses the Metropolis algorithm as distinguished from
maximum penalized likelihood.  These differences are discussed in
detail in the next section. The scheme reported in \citet{david} is a
generic algorithm that does not involve the assumption of isotropy; a
simple anisotropic distribution function that depends on both energy
and the line-of-sight component of the angular momentum is considered.
However, in this report, we solve the simpler case of the isotropic
distribution function. (We buttress our analysis with a check of
consistency of the data with isotropy).  The algorithm in
\citet{david}, which uses discrete data, is discussed below.

\subsection{A generic non-parametric method}
Let a stellar system be represented by an equilibrium distribution
function (DF), $f$, which is a function of energy $E$ and angular
momentum, $L$.  One way of projecting $f$ into the observable space is
by first projecting it on the plane of the sky and then along the line
of sight velocity axis in the phase space. The resulting projected DF
will then be dependent on the plane of sky coordinates and also on the
LOS velocity coordinate. This projected DF, say $\nu_p$, is related to
$f$ via the integral equation
\begin{equation}
\nu_p(x,y,v_p)=\int{dz\iint{dv_xdv_xf(E,L)}},
\label{eq:integral}
\end{equation}
where the LOS is along the $z$-direction while the plane of the sky is
scanned by the $x$-$y$ plane. If the stellar system being studied
exhibits spherical geometry, then the last equation becomes
\begin{equation}
\nu_p(r_p,v_p)=\int{dz\iint{dv_xdv_yf[v_r^2+v_t^2+2\Phi(r),L]}}. 
\label{eq:proj}
\end{equation}
Here $r$ is the spherical radius and $r_p$ the cylindrical radius
on the plane of the sky, so that $r_p^2=x^2+y^2=r^2-z^2$; $v_r$ and
$v_t$ are components parallel and tangential to the radius
${\bf{r}}$; $\Phi(r)$ is the gravitational potential of this system. To
simplify matters we could choose the $y$-axis to lie completely in the
plane containing the radius vector ${\bf{r}}$ and the LOS. This implies
that $v_r^2=(v_y\sin\theta)^2+(v_z\cos\theta)^2$, $\theta$ being the
polar angle, i.e., $\cos\theta=z/r$.

A non-parametric solution determines the potential and the
distribution function by searching for a pair of functions $f$ and $\Phi$
that provide the best-fit to the complete velocity data. The first
step is therefore to identify a way by which a trial DF could be
projected into observable space, according to eqn.~\ref{eq:proj}.  An
easy way to do this would be to approximate $f(E,L)$ as a 2-D
histogram.

Thus $f$ is approximated to be a constant over any integral cell
defined around a pair of $E$ and $L$ values. If, of course, the
assumption of isotropy is invoked, the integral space becomes one
dimensional so that $f$ is a function of $E$ only. $f$ is then
approximated to be a constant over any energy bin in this integral
space. The contribution to the projected distribution function from
each such energy bin is given by
\begin{equation}
\nu_p^{\rm cell}(r_p,v_p)=\int{dz\, A(r,r_p,v_p)},
\end{equation}
where $A(r,r_p,v_p)=\iint{dv_xdv_y}$ is the area that the energy
integral bin occupies in the $(v_x,v_y)$ space. The total projected 
distribution function is a simple sum over all the energy bins, i.e.,
\begin{equation}
\nu_p(r_p,v_p)=\sum_{i}f_i\nu_{pi}^{\rm cell}(r_p,v_p).
\label{eq:invert}
\end{equation}

To get the total distribution function $f$ from the projected DF, 
eqn.~\ref{eq:invert} has to be inverted, (eqn.~\ref{eq:invert} can be treated
as a set of linear equations). If $\nu_p(r_p,v_p)$ is a continuous function
of $r_p$ and $v_p$, the inversion is straightforward but if the data set
consists of a large number of discrete data points, then the projected DF
can be estimated by binning the data with respect to $r_p$ and $v_p$.

The probability that the observed kinematical data was drawn from 
$f(E)$ could be measured by the likelihood function ($L$) that 
can be defined as the product of all the projected distribution functions,
each obtained for a pair of $(r_p,v_p)$
\begin{equation}
\log(L) = \sum_{\rm{i=1}}^{\rm{N}}\nu_{pi},
\end{equation}
where $N$ is the total number of pairs of phase space points in the
observed data set and $\nu_{pi}$ is the projected distribution
function corresponding to the $i^{\rm th}$ pair of apparent position
and velocity in the data set. The optimization can be carried out in
the presence of linear constraints on the total number of stars,
presented as a penalty function.  Thus in
\citet{david}, the maximum penalized likelihood was employed.
 
\subsection{Unique features of our algorithm} 
Our method is essentially an isotropic version of the above,
but there are a few important differences between the two algorithms, 
as discussed below. 
\begin{enumerate}
\item We derive the potential from a discretized density
distribution.  The mass (including any dark matter) is assumed to be
in spherical shells with density decreasing outwards but otherwise free-form.
\item As discussed in the last section, in \citet{david}, a maximum
penalized likelihood was suggested to find out the distribution
function that the observed data was most likely to have been drawn
from.  When the likelihood is highly non-linear, it cannot be
optimized by linear constraints or even by quadratic programming,
(\citet{david96}).  In this case, the maxima in the likelihood can be
found by searching for it using brute force (\citet{davidsaha}).
However this method is rather unsatisfactory in terms of probability
of success and required CPU time. We use the Metropolis
algorithm, and attempt not only to recover the model with the maximum
likelihood, but to generate an {\it ensemble} of models distributed
according to the likelihood.  This ensemble will of course be
dominated by models close to the maximum likelihood model.  The
advantage of having such an ensemble is that we can estimate
uncertainties simply by measuring the spread of values across the
ensemble.  Moreover, we can use discrete data without ever having to
bin them.

\item Since we have both radial and proper motion data, we realized
that we could project the distribution function into observable space
in three ways. Firstly we could project the DF on the plane of the sky
and then project it along the line-of-sight velocity axis in the phase
space. This would give the projected DF, $\nu_p$ which will have
dependence on the plane of sky coordinates and the line of sight
velocity coordinate. We could also project the DF first on the plane
of the sky and then on the plane which is perpendicular to the line of
sight velocity axis in the velocity space. The resulting projected
distribution function is then dependent on the plane of sky
coordinates and also on the velocity coordinates perpendicular to the
line of sight velocity coordinate. We name such a projected
distribution function $\nu_{\perp}$. There is a small sample of stars
for which we know both radial and transverse velocities. This allows
us to project the distribution function along each of the three
velocity axes of the phase space, subsequent to the projection on the
plane of the sky.  The resulting projected distribution function will
be dependent on the apparent position on the sky and the radial and
tangential velocities.  We call such a projected distribution function
$\nu_{\rm 3D}$.

\item
We incorporate the effects of the errors in the velocity measurements
by convolving the calculated projected distribution functions with error
distributions that are assumed to be Gaussian.

\item
It is rightly argued that finding the data to be inconsistent with
anisotropy is a surer check of the assumption of an isotropic
distribution function than checking for the consistency of the data
with isotropy. But the former is a much harder test to execute. Hence
we carry out our analysis by starting with the assumption of isotropy
and at the end of analysis, perform a goodness-of-fit test to check
for the validity of the assumption for the observed data sets.
\end{enumerate}

\subsection{The Metropolis algorithm}
\label{sec:metropolis}
The algorithm used in our code to identify the distribution of
likelihoods is the Metropolis algorithm, \citep{metrop}.  The
likelihood function $L$ is a function of the potential and of the
distribution function histograms, i.e., $L$ is a function in
$n$-dimensional space, where $n$ is the number of energy bins in the
DF histogram times the number of mass shells. Let us represent the
likelihood as $L = L({\bf x})$ where ${\bf x}\equiv(\rho,f)$.

The code samples the function $L({\bf x})$ through a series of
iterations $L({\bf x}_n)$. Two consecutive iterations are related by a
transition probability $p({\bf x}_n\rightarrow{\bf x}_{n+1})$.  The
transition probabilities are chosen as
\begin{equation}
p({\bf x}\rightarrow{\bf x'}) =
{\rm max}\left( {L({\bf x'})\over L({\bf x})},1\right),
\end{equation}
although in principle any choice satisfying detailed balance
\begin{equation}
L({\bf x'})p({\bf x'}\rightarrow{\bf x}) =
L({\bf x})p({\bf x}\rightarrow{\bf x'})
\end{equation}
would serve.

\begin{figure}
\plotone{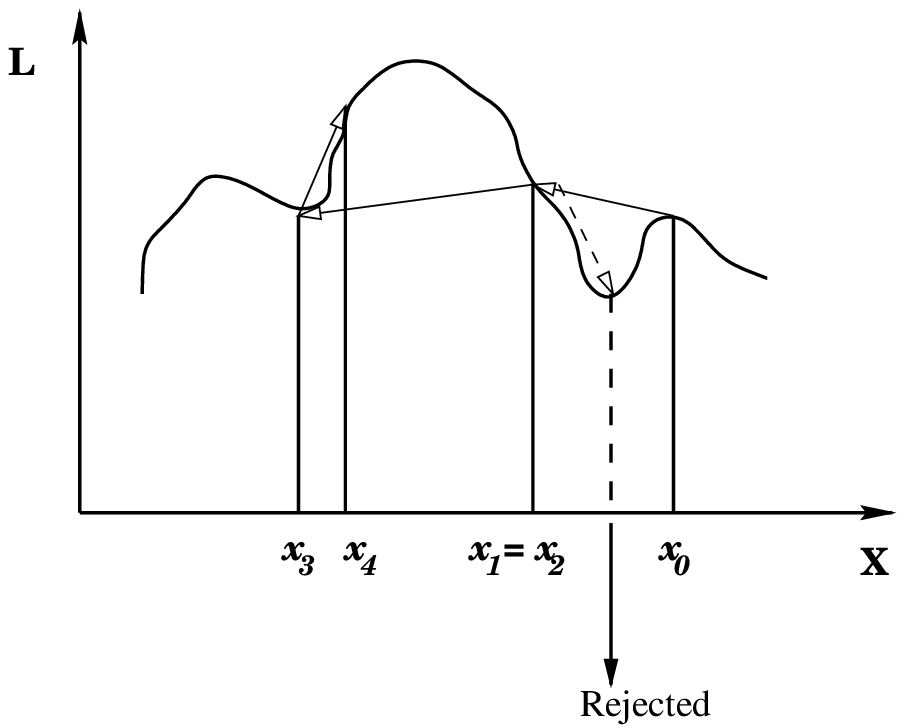}
\end{figure}

The algorithm is schematically represented in
Figure~\ref{fig:metrofig}.  At each iteration, the code tries a new
${\bf x}$, i.e., new $\rho$ and $f$.  These histograms are always
non-negative and monotonic, and with unit normalization for $f$, but
otherwise arbitrary.  The trial ${\bf x}$ is accepted or rejected with
the transition probability; in case of rejection the code keeps the
old ${\bf x}$ for the next iteration.

In its early phase Metropolis searches out the region around the
maxima in the likelihood and then the more ``equilibrium'' phase sets
in, when Metropolis wanders about in the region of the
multi-dimensional energy-density space, where the likelihood is close
to maximal. The extent of this wandering provides the uncertainties
in the mass estimates.

As the iterations proceed, the likelihood is monitored.  Initially,
there is trend towards increasing likelihood.  When this trend
disappears, Metropolis is in the equilibrium phase.  In practice we
waited till iteration $k$ such that $L({\bf x}_k)<L({\bf x}_{k-799})$,
and then took $\{{\bf x}_{k-799},\ldots,{\bf x}_k\}$ as our
likelihood-weighted sample of $\rho$ and $f$. We recorded 
$\{L({\bf x}_{k-799}), ..., L({\bf x}_{k})\}$. The mass in the innermost
radial bin, ($M_{\rm in}$) was calculated at iteration numbers $k-799$, ..., $k$,
from the likelihood-weighted $\rho$ obtained at the end of each iteration.
Thus a sample of these 800 values of $L$ and $M_{\rm in}$ were recorded.

For the test problem (Section \ref{sec:plum} below) we compared the
likelihood-weighted $f$ and $\rho$ thus generated, with $f$ and $\rho$ corresponding to
a known model. 

For real data, we added an extra precaution. We generated samples of
800 values of $L$ and $M_{\rm in}$ from 10 different initial
configurations.  To each sample we applied a goodness-of-fit test (see
below); the samples that passed that test were concatenated into a
master sample.  The item of interest from this master sample is of
course the mass in the innermost bin, histograms of which are displayed
later in this paper.  The median of such a histogram is quoted as the
mass estimate for the black hole in our Galaxy while the 68\%
confidence limits are given by the $16\%$ and $84\%$ levels.

The Metropolis algorithm must not be confused with the related
technique of simulated annealing.  Simulated annealing could be used
to locate the global maximum of $L({\bf x})$, but that is insufficient
for our purposes because we are interested not just in the
most-probable black hole mass but in its uncertainty.

\subsection{Goodness of fit}
\label{sec:isotropy}

To check our assumptions (especially isotropy of the velocity
distribution) and the fitting procedure, it is necessary to have
a goodness-of-fit test.

The first step in a goodness-of-fit test is to choose a statistic, or
measure of goodness-of-fit. There are many possibilities, but the
obvious one is the likelihood $L$.

The second step is to compute the statistic for the real data, and
compare its value with the expected distribution of that statistic
from the model. If 99\% of synthetic data that is drawn from the
model, fit better than the actual data then the model is rejected at
99\% confidence, and so on.

To implement such a test, we took $f$ from the last iteration of
Metropolis and generated 100 synthetic kinematical data sets of the same size as
the real data set, and evaluated the statistic (i.e., $L$) for each of
these. We then computed the fraction of the synthetic data sets that fit
{\it less well\/} than the real data.  We call this fraction
$P$,\footnote{This is sometimes called the $p$-value of a statistic.}
and quote it as a percentage.  Low $P$ implies that the fit must be rejected.

If we had been doing least-squares, the above prescription would
reduce to a $\chi^2$ test, since $\chi^2=\frac12\ln L$ in
least-squares.  The expected distribution of $\chi^2$ is
model-independent (as is the expected distribution of the KS
statistic), so there is no need to generate synthetic data when using
it.

For acceptable fits, $P$ is not uniformly distributed between 0 and
100\%, but tends to be on the high side.  This is because the fitting
procedure has already in some sense optimized $P$.  In least-squares,
this effect is corrected for by subtracting the number of fitted
parameters from $\chi^2$; we do not know if this correction can be
generalized.

The value of $P$ has been pictorially depicted in Figures
\ref{fig:barerr} and \ref{fig:barnoerr}; for a given initial model,
$P$ is measured and plotted in a stacked bar diagram, with the three
different columns corresponding to the three different observed
kinematical data-sets used in the work.

\section{Implementation details of our algorithm}
\label{sec:details}

\subsection{Calculation of the potential}
The mass distribution is a series of concentric spherical shells
each with different densities.  The potential
\begin{equation}
\Phi = \frac{1}{r}\int_{0}^{r}{\rho(r'){r'}^{2}dr'} +
\int_{r}^{\infty}{dr'\rho(r')r'}.
\end{equation}
can then be expressed as follows.  If $r$ falls within the $m$-th
shell (of $n$ shells in all) then
\begin{equation}
\Phi = \frac{1}{r}\sum_{\rm{i=1}}^{\rm{m-1}}
{\rho_i(i^2-i+\frac{1}{3})\delta^3} +
\sum_{\rm i=m+1}^{n}{\rho_i(i-\frac{1}{2})\delta^2}+
\rho_m\left[-\frac{r^2}{6}+\frac{m^2\delta^2}{2}-
\frac{(m-1)^3\delta^3}{3r}\right],
\end{equation}
and if $r$ is outside all the shells then
\begin{equation}
\Phi = \frac{1}{r}\sum_{\rm i=1}^{n}{\rho_i(i^2-i+\frac{1}{3})\delta^3}.
\end{equation}
Here $\delta$ is the thickness of each shell.

We constrain $\rho$ to be non-increasing as $r$ increases, but apart from
that the density shells can take any positive values.

It needs to be emphasized here that we seek the mass of the black hole
in the Galactic center as the mass that is included inside a pre-fixed
radius, $r_{\rm in}$. This is the projected radius at which the
innermost velocity data measurement is reported, in the used data
sets.  In the radial velocity measurement by \citet{genzel1}, $r_{\rm
in}\approx0.046$pc while in the proper motion work by \citet{ghez}, it
is about 0.0044pc.  At $r<r_{\rm in}$, the code will spread the mass
uniformly. Hence the best that we can do with the available data is to
ask for the mass inner to the radius at which the innermost
measurement has been recorded. As pointed out in \citet{ghez}, the
mass estimate could be totally attributed to a central black hole only
if resolution improves to the extent that $r_{\rm in}$ corresponds to
the Schwarzschild radius of the black hole. This is merely wishful at
this stage as the minimum radius at which a measurement is recorded in
\citet{ghez} is about 40,000 times less than the Schwarzschild radius
of a $2.6\times10^6{\rm M}_\odot$ black hole.

\subsection{The projected distribution functions} 
\label{seec:projected}

In the coordinate system that we adopt, the plane
of the sky is scanned by the $(x,y)$ plane while the line-of-sight
is along the $z$ direction.

The projected distribution function $\nu_p$ is given as
\begin{equation}
{\nu_{p}(r_p, v_z)} = \int{dz}\iint{f(E)dv_xdv_y}.
\end{equation}
As discussed before, $\nu_{\perp}$ is in general dependent on the
apparent position and the velocity coordinates that are perpendicular
to the line of sight velocity axis in velocity space, i.e., on $v_x\;\&\;v_y$. 
If we assume axisymmetry in velocity space, on the plane
perpendicular to the $\bf{v_p}$ axis, then we can represent
$\nu_{\perp}$ as the equation below, (where $v_{\mu}=\sqrt{v_x^2+v_y^2}$),
\begin{equation}
\nu_{\perp}(r_p, \sqrt{v_x^2+v_y^2}) = 
\int{dz}\int{f(E)dv_z}, 
\end{equation}
and $\nu_{\rm 3D}$ is given as
\begin{equation}
\nu_{\rm 3D}(r_p, v_\mu, v_z) = \int{dz\quad f(E)}. 
\end{equation}

The bounds on individual phase space
coordinates that correspond to each such energy bin are
identified. These bounds are then used as limits for the projection
integrals that relate the projected DF for the $i^{\rm th}$ energy bin
to the constant value (say $f_i$) of the total DF over this energy bin. 
It must be noted that the bounds on the phase space variables may depend
on how the projection is executed. 

Let the $i^{\rm th}$ energy bin run from energy $=E_{1i}$ to $E_{2i}$.
Since $r^2=r_p^2+z^2$, the minimum value that $r$ could take over the
current energy bin, (for a given value of the observed apparent
position) is $r_p^2$, obtained by setting $z$ to zero.  Let the
maximum value that $r$ could have in this energy bin be $r_{2i}$.  Now
$r$ is maximum for a given energy when the kinetic energy is a
minimum.  For the projection of the distribution function along the
vector $\bf{v_p}$, $v_p$ is a measured quantity. Hence the minimum
kinetic energy is obtained by setting $v_{\mu}$ to zero. This implies
that when $f$ is projected along the LOS velocity, the upper limit on
$r$ for a specified energy bin is: $r_{2i}$, where $r_{2i}$ is the
root of
\begin{equation}
E_{2i} = \frac{1}{2}v_p^2+\Phi(r_{2i}). 
\end{equation}
Similarly when $f$ is projected perpendicular to the LOS velocity,
$v_{\mu}$ is a measured quantity. Therefore, minimizing kinetic energy
implies setting $v_p$ to zero. Thus $r_{2i}$ in this case is defined
as
\begin{equation}
E_{2i} = \frac{1}{2}v_{\mu}^2+\Phi(r_{2i}) .
\end{equation}
The roots are found in our algorithm with one of the standard
root-finding routines.

For a given $r$, the speed $v=\sqrt{v_p^2+v_{\mu}^2}$ is extremal
when the energy is either a maximum or a minimum.
The limits on speed $v$ ($v_1$ being the lower
and $v_2$ the upper limit) are such that
\begin{equation}
\frac{1}{2}v(r)_{ji}^2=E_{ji}-\Phi(r), \quad\hbox{where $j=1,2$}. 
\end{equation}
Using the transformations
\begin{equation}
r_p^2 =r^2-z^2, \quad v_{\mu}^2 =v^2-v_p^2,
\end{equation}
we can write the integrals for the projected DFs over the current energy bin as
\begin{eqnarray}
\nu_{i\perp}(r_p,v_{\mu}) = f(E_i)\:\int_{r_p}^{r_{2i}(v_{\mu})}\frac{r[\sqrt{v_{2i}^2(r)-v_{\mu}^2}-\sqrt{v_{1i}^2(r)-v_{\mu}^2}]}{\sqrt{r^2-r_p^2}}dr\\
\nu_{ip}(r_p,v_p) = f(E_i)\:\int_{r_p}^{r_{2i}(v_p)}\frac{r[v_{2i}^2(r)-v_{1i}^2(r)]}{\sqrt{r^2-r_p^2}}dr\\
\nu_{i3}(r_p, v_p, v_\mu) = f(E_i)\:\int_{r_p}^{r_{2i}(v_p,v_{\mu})}\frac{r}{\sqrt{r^2-r_p^2}}dr
\end{eqnarray}

To get the projected DFs, we need to sum over all the energy bins.
(We constrain $f$ to be non-increasing as $E$ increases, i.e., $f$ is
greatest for the most tightly bound orbits.) In this whole scheme
$r_p\;\&\;v_p$ and $r_p\;\&\;v_{\mu}$ are read from the supplied data
set. The sky is seen at the different data points, each data point
being independent. Thus the probability that the positions and
velocities in the data set have been drawn from the currently chosen
distribution function will be measured by the product of the values of
the projected DF obtained with the apparent position and velocity
values from the whole data set. This product is then defined to be our
likelihood function, $L$.  Thus
\begin{equation}
\log L = \sum_{\rm data\ pts}{\log \nu_{k}} \quad
\hbox{where $k\equiv$ ``$\perp$'' or ``$p$'' or ``3D''}	\\
\end{equation}
The aim of this scheme is to start with an arbitrary pair of
density and distribution function histograms and sample according to
the likelihood function.

\section{Tests of the Method}\label{sec:plum}

We tested our code on a Plummer model, with simulated data sets of
similar size to the real data sets.

From the Plummer distribution function at the Plummer potential
\citep{bible}, we generated a data set containing 100 proper motion
values, another containing 200 radial velocity values and a third with
30 values of both radial and transverse velocities. These numbers
correspond roughly correspond to the populations in the observed data sets that
we use.

The Plummer model is the polytrope with a polytropic index
of 5. The distribution function for a general polytrope with index $n$
is given as:-
\begin{equation}
f = F(-E)^{(n-\frac{3}{2})}.
\end{equation}
In our work we decided to set the total mass in the Plummer model to unity.

The aim is to start with an arbitrarily chosen pair of DF and density
histograms, use them in the code that was described above, along with
the data drawn from the Plummer distribution function. The forms of
the density and distribution function profiles corresponding to the
convergence of the Metropolis algorithm maximum value of the
likelihood function were then compared with the Plummer density and
distribution function forms.

The comparison between the final distribution function and density
profiles thrown up by our code and the actual Plummer density and
distribution function profiles are presented below. The initial distribution
function and density profile for which Figure~\ref{fig:plumres} was obtained
were chosen as
\begin{equation}
f  = F(-E)^6, \qquad \rho = (1 + r)^{-0.9}.
\end{equation}

\begin{figure}
\plotone{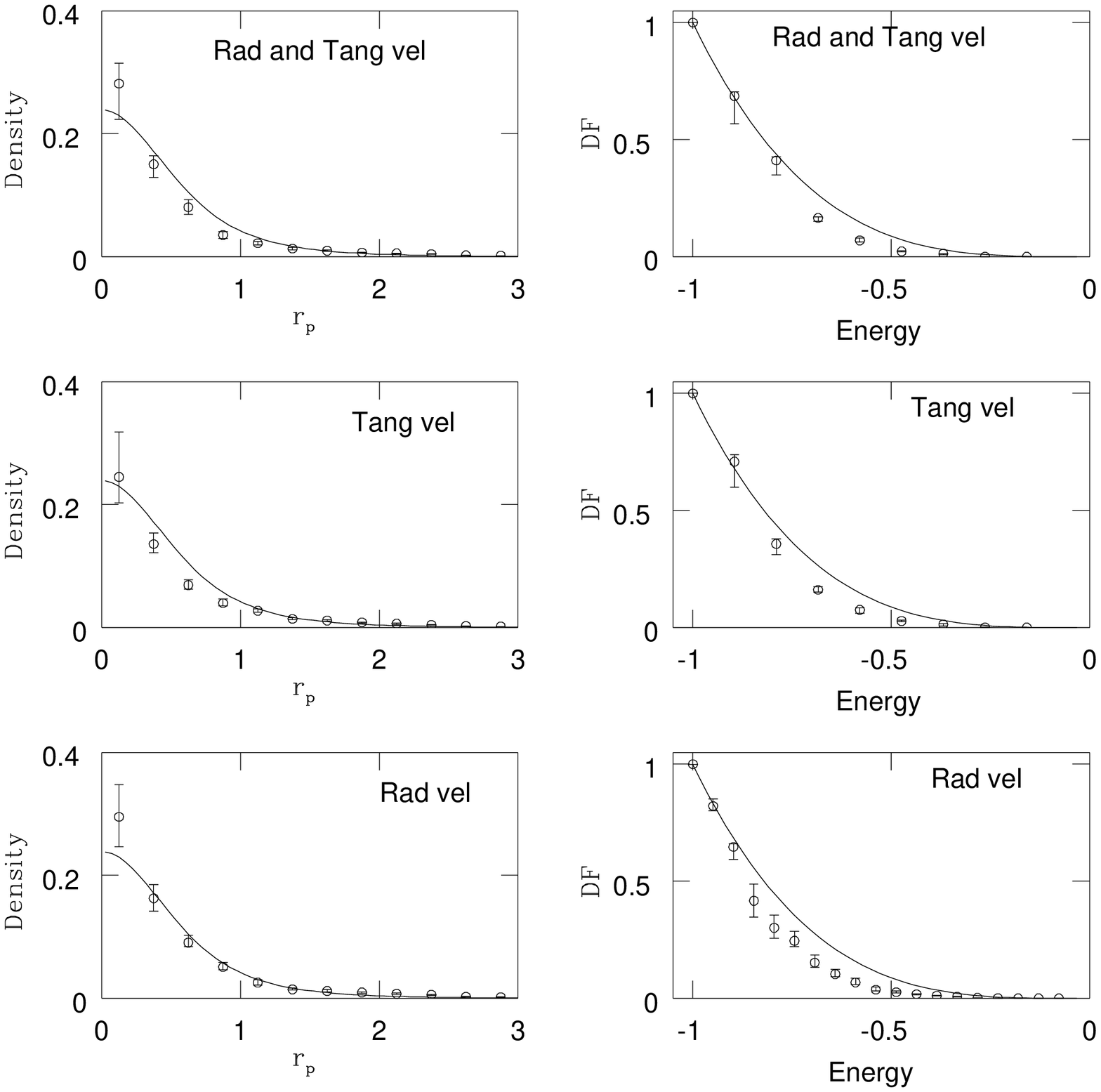}
\end{figure}

Figure~\ref{fig:plumres} shows that the comparison is quite good
between the actual Plummer model and the profiles suggested by our
algorithm. Encouraging comparisons were obtained even when the initial
density profiles were of the Dehnen model type and the initial
distribution function had a power-law dependence on $1 + (-E)^2$.

\section{Results}

\subsection{Initial models used with observed data}
\label{sec:initial}

The initial density profiles that we chose to input into Metropolis,
while working with the observed data sets, correspond to the Dehnen 
models,
\begin{equation}
\rho = \frac{(1+\alpha)}{4\pi}r^{\alpha-2}(1+r)^{-\alpha-2}\label{eq:dehnen}
\end{equation}
in the notation of \citet{saha93}.
The density profiles in eqn.~\ref{eq:dehnen} can represent different mass
distributions depending on the value of the parameter $\alpha$. For example,
$\alpha=-1$ implies a point mass while $\alpha=0$ represents a Jaffe
model---see \citet{binmer} for details.
In our work $\alpha$ was varied in the range $[-0.1,-1.0]$.

As mentioned before, the distribution function in this work has been
modeled as a function of energy only. The initial form of the
distribution function was chosen to be a power law.
\begin{equation}
f \propto (-E)^\beta \label{eq:expodf}
\end{equation}
with $\beta$ either 1.0 or 2.0.

\subsection{Distribution functions and density profiles}

Figure~\ref{fig:dfden} shows the density profiles and distribution
functions generated by Metropolis to fit the three data sets.  This
particular run started with a relatively steep DF and density.  We find
that $f$ does not vary much over the Metropolis sample---there is a strong
tendency for $f$ to stay close to the initial state.  For this
reason we decided to collate the results from 10 initial states.

\begin{figure}
\plotone{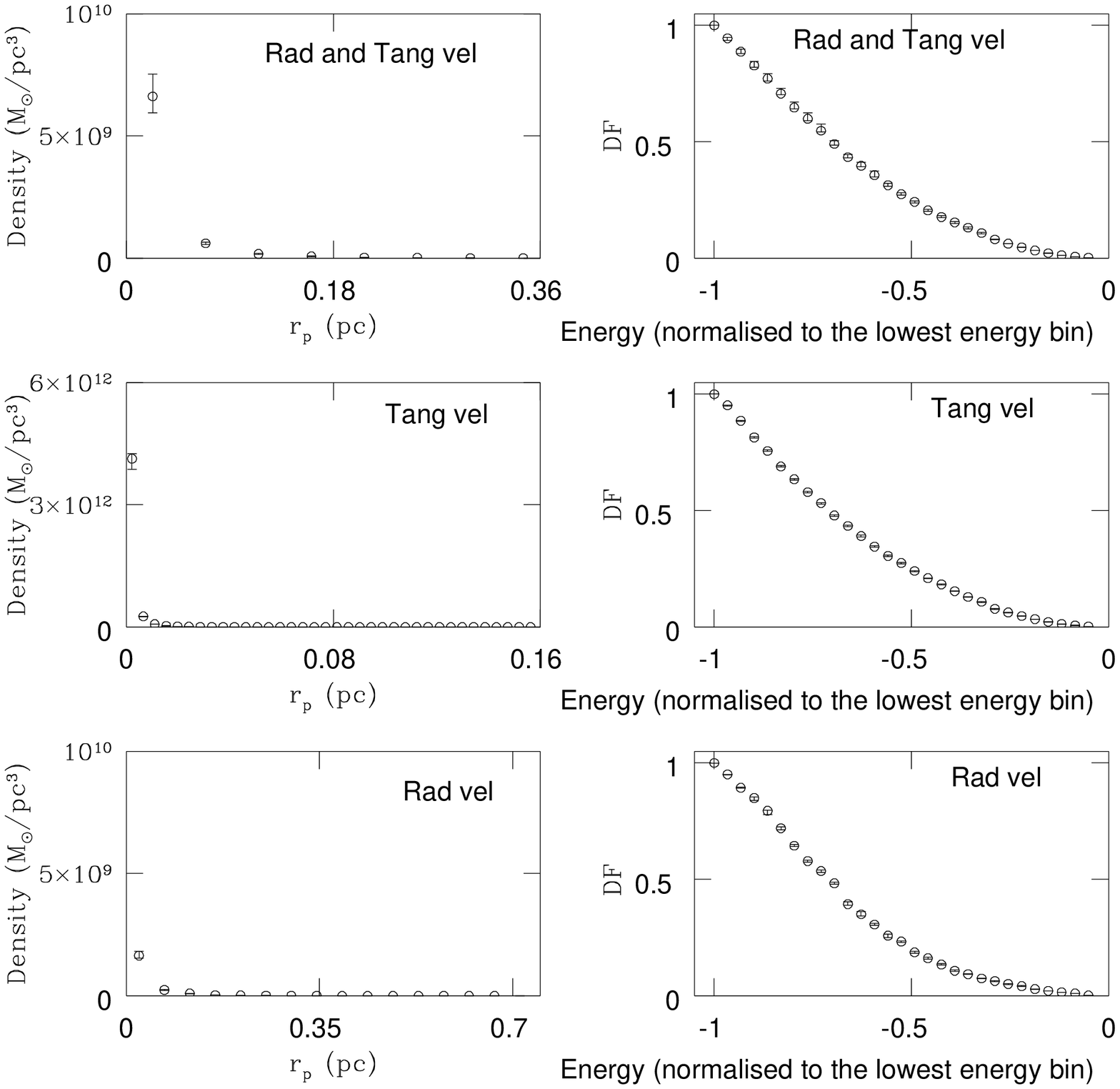}
\end{figure}

\subsection{Working with proper motion data}
\label{sec:ghez}

Our toy data obtained from the Plummer model gave us confidence to
use observed proper motion data presented in \citet{ghez}
to obtain an estimate of the mass of the central black hole in our
Galaxy.

The proper motion values cited in \citet{ghez} are accompanied by
observational errors. While a series of runs were carried out, without
taking into account the reported errors in the transverse velocities,
another series of runs were done in which the errors were convolved
with the projected distribution functions obtained with the
corresponding pair of transverse velocity and apparent position.  The
error distribution was assumed to be a Gaussian.  The initial models
for both types of runs were the same, (Sec.\ref{sec:initial}).

The histogram of the likelihood weighted estimate of the mass within
$r_{\rm in}=0.0044$pc is shown in Figure~\ref{fig:histmu}.  The
histograms were constructed for the sample of mass estimates in the
way described in Sec.\ref{sec:metropolis}, with those models that
passed our goodness-of-fit test (Sec.\ref{sec:isotropy}). For both the 
error-convolved and error-ignored runs, the initial models with the
less steep distribution functions, passed the test satisfactorily
while the initial DFs $\propto (-E)^2$ gave very poor values of the
parameter $P$ defined in Sec.\ref{sec:isotropy}.  The dependence of
$P$ on the initial models in the runs done with the proper motion data
are represented in Figure~\ref{fig:barerr} and
Figure~\ref{fig:barnoerr}.

From the error-convolved distribution of mass estimates shown in
Figure~\ref{fig:histmu}, we find a median of $2.0\times10^6{\rm
M}_{\odot}$.  At the lower 16\% of the distribution, the mass is
$1.3\times10^6{\rm M}_{\odot}$ while at the upper 84\% the mass is
$2.7\times10^6{\rm M}_{\odot}$.

\begin{figure}
\plotone{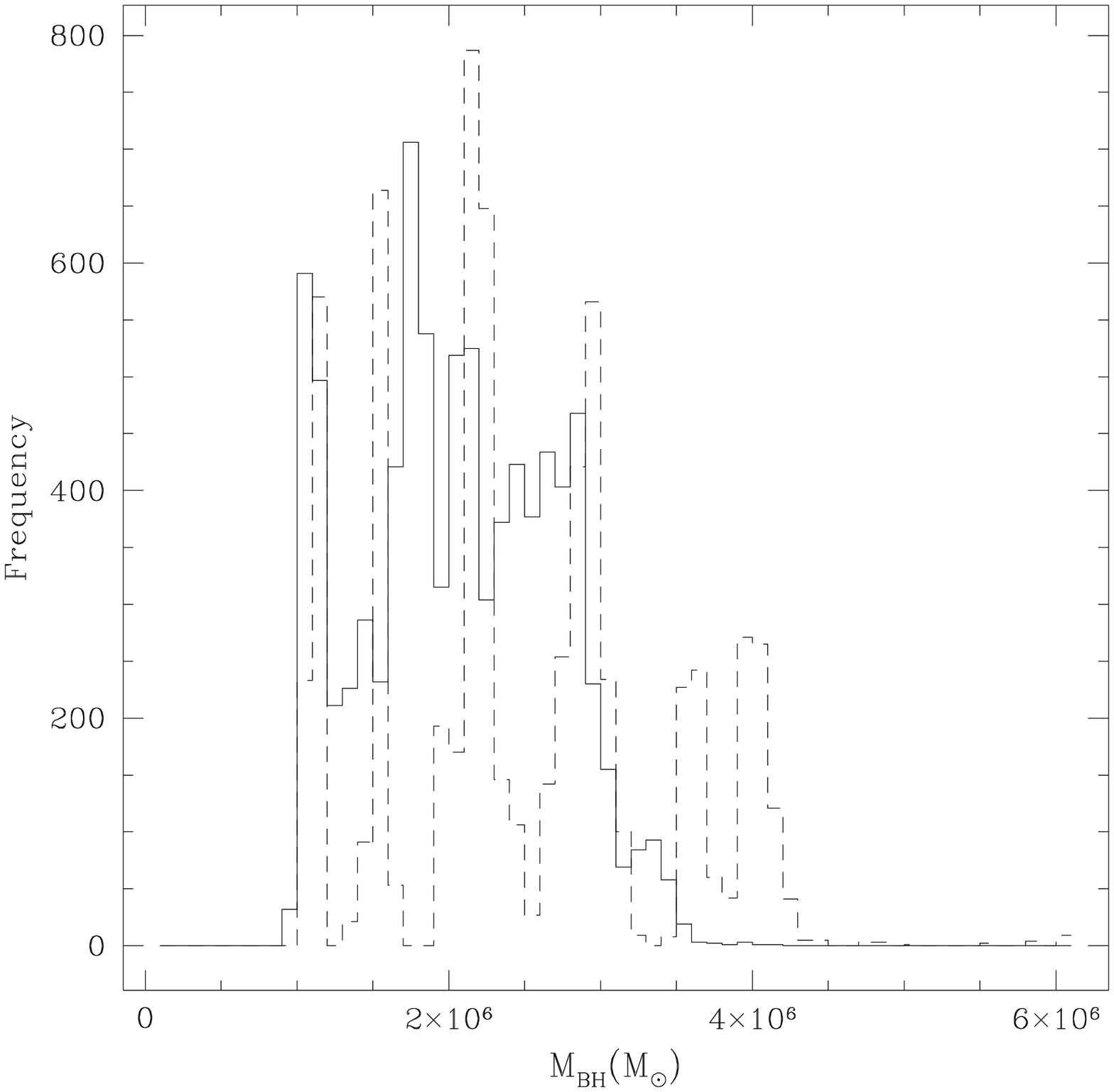}
\end{figure}

\subsection{Working with the radial velocity data}

We have used $204$ of the radial velocity data and the errors in the
same, reported in \citet{gerhard} to obtain another mass estimate of the black
hole at the center of the Galaxy. For this data set, $r_{\rm in}$ is 0.046pc.

\begin{figure}
\plotone{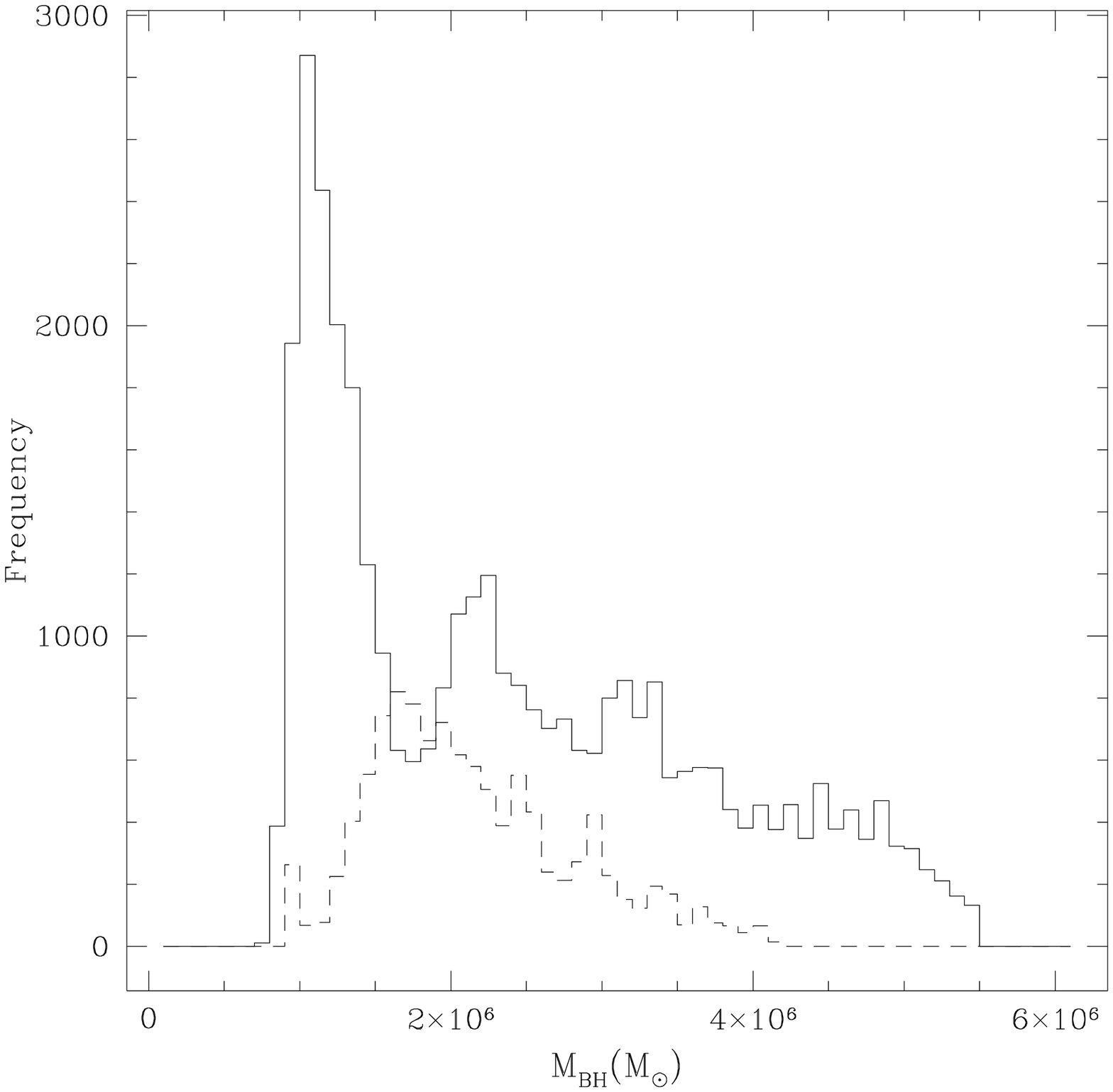}
\end{figure}

The initial distribution function and density profiles used in
conjunction with the radial velocity data were as mentioned in
eqn.~\ref{eq:expodf} and eqn.~\ref{eq:dehnen} respectively. For the
runs in which observational errors were incorporated, the distribution
functions were convolved with the errors, the error distribution being
assumed to be Gaussian.  The histograms representing the values of
likelihood weighted mass, are shown in Fig~\ref{fig:histrad}.  The
histograms were obtained using the same scheme described in
Sec.\ref{sec:metropolis}. The histograms were constructed with only
those initial models, which passed the goodness-of-fit test
(Sec.\ref{sec:isotropy}) satisfactorily. All the considered
initial models pass the goodness-of-fit test quite well when
observational errors are considered, but only half of the initial
models do so when errors are ignored. Hence the sample of mass
estimates for the error-convolved runs is much bigger than the same
when errors are ignored. This also explains the longer tail of the
error-convolved histogram.

We find that in the case of the error-convolved runs done with the
radial velocity data set, $P$ (defined in Sec.\ref{sec:isotropy}) is
smaller for the runs done with the steeper initial distribution
function ($\propto(-E)^2$) than with the flatter ones ($\propto(-E)$),
but is never zero.  The variation in $P$ with initial model is shown
in Figure~\ref{fig:barerr} and Figure~\ref{fig:barnoerr}.

From the error-convolved histogram in Figure~\ref{fig:histrad}, we obtain
a median of $2.2\times10^6{\rm M}_{\odot}$ while the mass estimates at
the 16\% and the 84\% levels are $1.2\times10^6{\rm M}_{\odot}$ and
$3.8\times10^6{\rm M}_{\odot}$ respectively.

\begin{figure}
\plotone{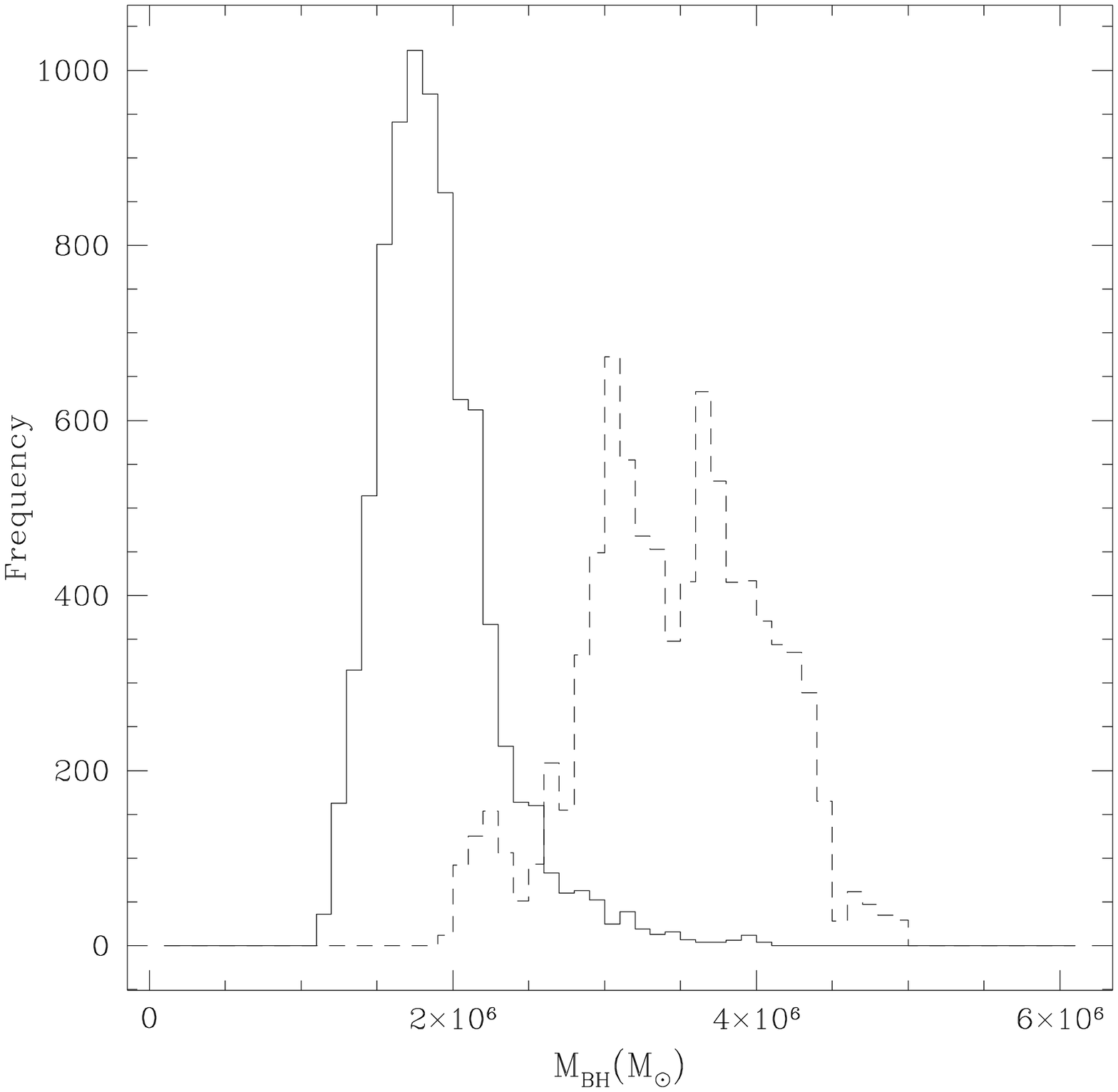}
\end{figure}

\begin{figure}
\plotone{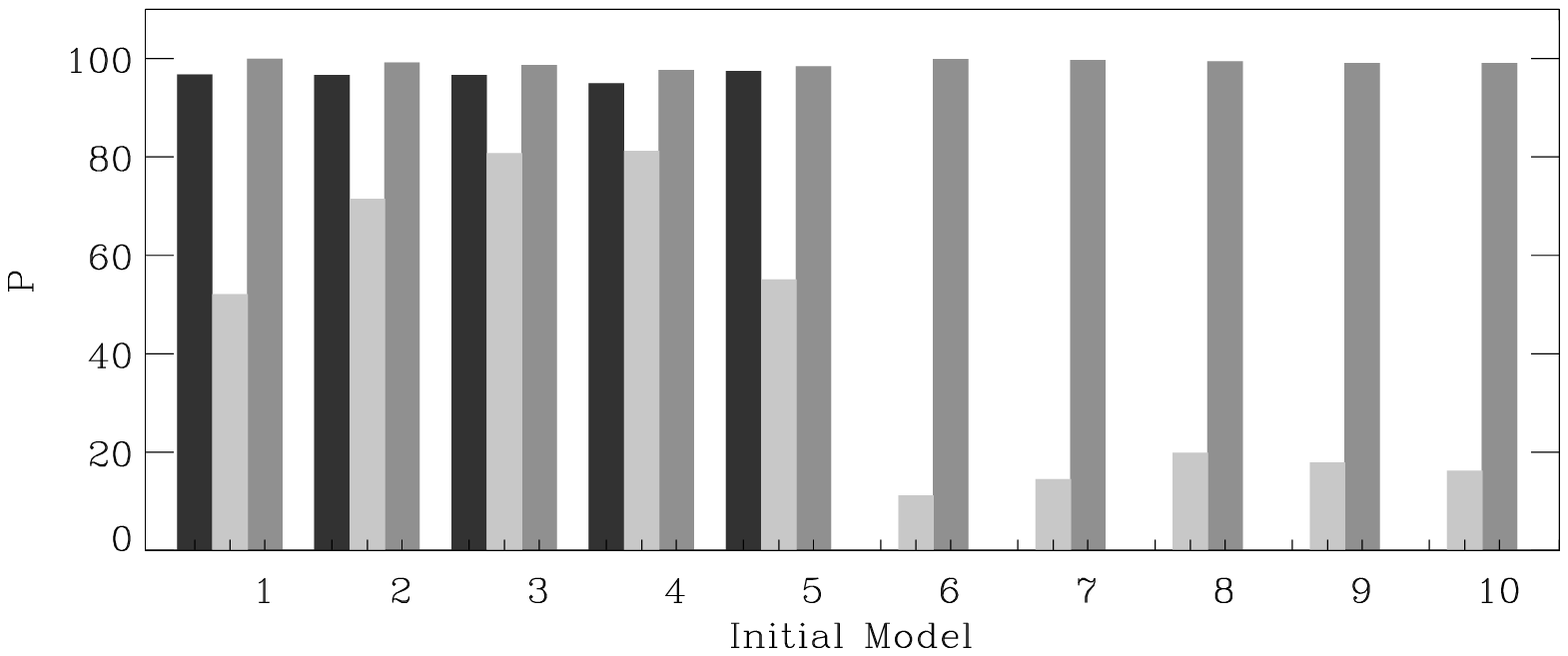}
\end{figure}

\begin{figure}
\plotone{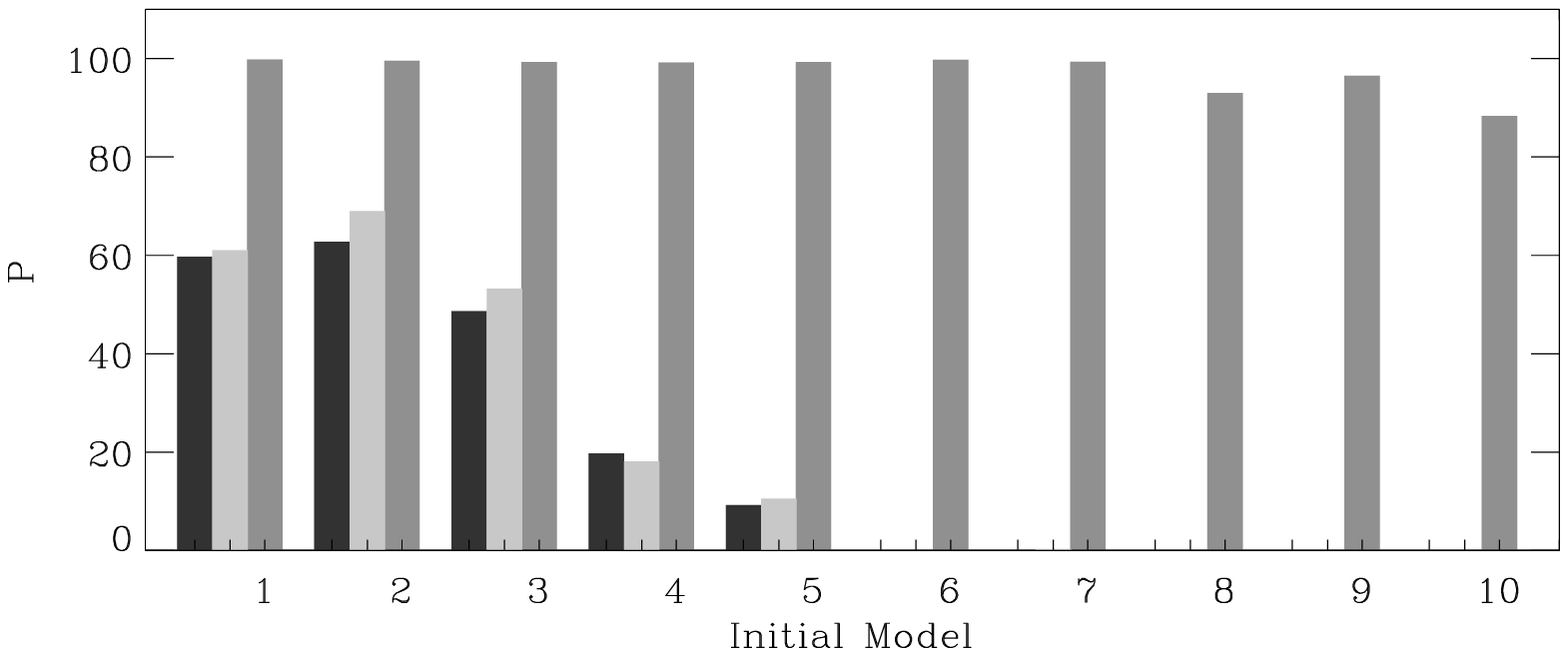}
\end{figure}

\subsection{Working with the 3-D motion data}

In \citet{gerhard}, $32$ stars were reported with their proper motion
and radial velocities as well as the errors in the same.  For this
data set, $r_{\rm in}$ was 0.046pc.  Histograms were constructed with
the same initial models as mentioned in Sec.\ref{sec:initial} for runs
done both by convolving the projected distribution functions with
errors as well as by ignoring the observational errors. These
histograms can be seen in Figure~\ref{fig:histthr}. All the considered
initial models were found to pass the goodness-of-fit test,
(Sec.\ref{sec:isotropy}) for both types of runs, (with and without
errors).

From the error-convolved histogram in Figure~\ref{fig:histthr}, we obtain a
median of $1.8\times10^6{\rm M}_{\odot}$ while the mass estimates at
the 16\% and the 84\% levels are $1.5\times10^6{\rm M}_{\odot}$ and
$2.2\times10^6{\rm M}_{\odot}$ respectively.

\subsection{Results explained}
Some of the results presented above have been qualitatively discussed
below.

The distribution of the mass estimates has a non-zero width for two
reasons.
\begin{itemize}
\item the mass estimate sample is recorded while Metropolis is
wandering around the maximal region of the likelihood function during
the ``equilibrium phase'' though it makes occasional journeys to
lower likelihoods too.
\item the mass estimate sample is composed of results obtained from
different initial models. Though the initial model is not expected to 
dictate the convergence to the true distribution function (from which the
observed data is drawn), the limited number of data points does limit
the independence of the results from the initial model.
\end{itemize}

The results obtained with any initial model could fail the
goodness-of-fit test if there is any systematic error (or bias) in the
code (or the data) or if indeed the observed data is not consistent
with isotropy or if Metropolis was unable to seek out the optimal fit
to the observed data. If the test is passed, it implies that the used
kinematical data agrees with isotropy. As described above, we carried
out this test with a variety of initial distribution function and
density profiles. We find that while the test is passed by some of the
models, (usually the models characterized by the relatively flatter
initial DF) when the radial or transverse velocity data is used, all
the models easily pass the goodness-of-fit test when the 3-D motion
information is used. If indeed there was anything systematically
wrong, the analysis done with all the initial models would give
similar results during the check for isotropy. This leads us to
conclude that the reason for the failure of any initial model to pass
the goodness-of-fit test is not a systematic error.

The error-convolved histograms usually imply a smaller mass for the
black hole, for any given data set. The effect of errors in the
velocity measurements is equivalent to introduction of noise, which
implies greater energy for a star at a given phase space location than
when the noise is absent. This in turn could (if uncorrected) imply a
higher enclosed mass. When the error distribution is convolved with
the projected distribution function, we compensate for this spurious
enhancement in mass. This effect of noise in the data is conjectured
to be increasingly overwhelmed with increasing number of used observational
data points. Hence we would expect in our results that the broken
histogram, (representing the distribution of mass estimates in absence
of the error convolution) is furthest from the solid histogram,
(representing the distribution of mass estimates obtained from the
error-convolved runs) when the mere $32$ data point strong 3-D
velocity data is used.

A steeper distribution function implies lower value of $f$ at most
energies than does a flatter distribution function. Hence the flatter
DF will imply higher likelihoods. Thus the likelihoods collected
during the final ``equilibrium'' phase of Metropolis, with the steeper
initial models, fall short of the simulated likelihoods. Since the
limitation in the number of data points is the chief factor
responsible for putting a brake on the mutual exclusivity of the
results and the initial models, all the considered models pass the
goodness-of-fit test with the bigger radial velocity sample during the
error-convolved runs than the smaller proper motion sample in
\citet{ghez}. However, when the number of data points goes up, the
``true'' fit to the observed data, as sought by Metropolis,
deteriorates since the maxima of $L$ gets progressively narrower with
increasing number of kinematical data points, (all of the considered
models pass the goodness-of-fit test when the much smaller 3-D motion
data set is used).  Thus there is a trade-off between the goodness of
fit achieved by Metropolis and the independence of the results from
the initial models.

It is also observed that a greater number of the initial models appear
to confirm consistency of the data with isotropy (via the
goodness-of-fit test), when the projected distribution functions are
error convolved. As explained above, the introduction of noise in the
data in the form of errors in the velocity measurement causes an
enhancement in the stellar energies. Thus for a given initial model,
noise pushes down the value of the likelihood. When this noise is
compensated for, (via convolution of the error distribution and the
projected DF), the likelihood function is expected to rise to its
correct value again, which may fall within the range of the simulated
likelihood sample.

\section{Summary}
We propose a new method for analyzing discrete measurements of stellar
radial velocities and/or proper motions and apply it to Galactic
center stellar dynamics, where there are currently $\sim 200$ radial
velocities, $\sim 100$ proper motions, and $\sim 30$ three-dimensional
velocities now available.  Both the distribution function and the
underlying potential are recovered non-parametrically.  This problem
was previously intractable because of the technical difficulty of
searching through a huge space of possible potentials and distribution
functions; we overcome this difficulty by applying the Metropolis
algorithm.

In this paper we consider the simplest case of a spherical isotropic
system.  We have designed a goodness-of-fit test to check whether the
data exclude this case, based on which we conclude that this
assumption is adequate under current data.

From proper motions reported by \citet{ghez}, we estimate
$2.0\pm{0.7}\times10^6{\rm M}_{\odot}$ within a radius of $0.0044$pc
when observational errors are incorporated into the analysis.  From
radial velocities reported by \citet{gerhard}, we estimate
$2.2^{+1.6}_{-1.0}\times10^6{\rm M}_{\odot}$ within $0.046$pc, and
from three-dimensional velocities reported again in \citet{gerhard},
the mass obtained is $1.8^{+0.4}_{-0.3}\times10^6{\rm M}_{\odot}$
within $0.046$pc.

\clearpage
\figcaption[dcps.fig1.ps]
{Schematic diagram to illustrate the Metropolis algorithm.
The code starts at ${\bf x}_0$, and proceeds to higher $L$ at ${\bf
x}_1$.  A trial ${\bf x}_2$ at lower $L$ is rejected, and so ${\bf
x}_1$ gets repeated as ${\bf x}_2$.  Next, the code goes to ${\bf
x}_3$, also at lower $L$, and then on to ${\bf x}_4$.  The code
wanders around near the global maximum of $L$, the sampling being
proportional to $L$.\label{fig:metrofig}}

\figcaption[dcps.fig2.ps]
{Plot to compare actual Plummer distribution function and
density profiles (in solid lines) with those obtained from Metropolis
runs done with artificial kinematical data drawn from the Plummer
model (shown as points with overlapping error bars).  The panels on
the left represent density plots while the panels on the right show
distribution functions. The two lower-most panels exhibit the
comparison when 200 artificial radial velocity data were used, the
panels in the second row correspond to the run done with 100
transverse velocity data and the uppermost panels correspond to a run done with 30
values of 3D velocities.\label{fig:plumres}}

\figcaption[dcps.fig3.ps]
{Distribution functions and density profiles obtained with the
three different kinematical data sets, starting from steep initial
$\rho$ and $f$.  The figure is analogous to Figure~\ref{fig:plumres},
except that the energy is normalized by the energy corresponding to
the lowest energy bin. The smallest radii at which a velocity
measurement was reported in the radial velocity, proper motion and 3-D
velocity data sets are 0.046pc, 0.0044pc and 0.046pc respectively.
\label{fig:dfden}}

\figcaption[dcps.fig4.ps]
{Mass estimates using proper motion data. The histograms
show the mass within the inner $0.0044$pc obtained from runs consistent
with isotropy.  The solid-line histogram is for error-convolved
runs while the broken-line histogram is for the error-ignored runs.
In this case, all the runs found to be consistent with isotropy
had a flat initial DF profile.\label{fig:histmu}}

\figcaption[dcps.fig5.ps]
{Mass estimates using radial velocity data.  Analogous to
Figure \ref{fig:histmu}, but for mass within $0.046$pc.  In this case,
all the error-convolved runs were found to be consistent with
isotropy, whereas only the error-ignored runs with flat initial DF
profiles were consistent with isotropy. Hence the different
normalizations of the histograms.\label{fig:histrad}}

\figcaption[dcps.fig6.ps]
{Mass estimates using 3D velocities.  Analogous to
Figures \ref{fig:histmu} and \ref{fig:histrad}, and showing mass within
$0.046$pc.  In this case all the runs were consistent with isotropy.
\label{fig:histthr}}

\figcaption[dcps.fig7.ps]
{Bar chart showing the results of the goodness-of-fit test
defined in Section \ref{sec:isotropy}, for error-convolved runs. The
dependence of the goodness-of-fit parameter, $P$, on the initial
models is shown. Bars represent the proper motion (black), radial
velocity (light grey) and 3-D motion (dark grey) data sets.  Models
$1-5$ represent flat DF profiles and density profiles that corresponds
to Dehnen models of increasing steepness, while models $6-10$
represent the same initial density functions and initial DF $\propto
(-E)^2$.\label{fig:barerr}}

\figcaption[dcps.fig8.ps]
{Analogous to Figure~\ref{fig:barerr}, but for error-ignored runs.  As
before, bars correspond to proper motion (black), radial velocity (light
grey) and 3-D motion (dark grey) data sets.\label{fig:barnoerr}}

\end{document}